\documentclass[12pt]{article}
\usepackage{graphicx}
\usepackage{amsmath}
\usepackage{amsfonts}
\usepackage{amssymb}

\begin{document}

\title{From a toy model to the double square root voting system}
\author{Wojciech S\l omczy\'{n}ski$^{a}$\\$^{a}$Institute of Mathematics, 
Jagiellonian University, \\ul. Reymonta 4, 30-059 Krak\'{o}w, Poland
\and Karol \.{Z}yczkowski$^{b,c}$\\$^{b}$Institute of Physics, 
Jagiellonian University, \\ul. Reymonta 4, 30-059 Krak\'{o}w, Poland \\
$^{c}$Center for Theoretical Physics, Polish Academy of Sciences, \\
Al. Lotnik\'{o}w 32/46, 02-668 Warszawa, Poland}
\date{October 9, 2007}
\maketitle

Abstract:

We investigate systems of indirect voting based on the law of Penrose, in
which each representative in the voting body receives the number of votes
(voting weight) proportional to the square root of the population he or she
represents. For a generic population distribution, the quota required for the
qualified majority can be set in such a way that the voting power of any state
is proportional to its weight. For a specific distribution of population the
optimal quota has to be computed numerically. We analyse a toy voting model
for which the optimal quota can be estimated analytically as a function of the
number of members of the voting body. This result, combined with the normal
approximation technique, allows us to design a simple, efficient, and flexible
voting system, which can be easily adopted for varying weights and number of players.\medskip

\medskip


\section{Introduction}

A game theory approach proved to be useful to analyse voting rules implemented
by various political or economic bodies. Since the pioneering contributions of
Lionel Penrose~(1946) who originated the mathematical theory of voting power
just after the World War II, this subject has been studied by a number of
researchers, see, e.g. Felsenthal and Machover (1998, 2004a) and references therein.

Although the current scientific literature contains several competing
definitions of voting indices, which quantitatively measure the voting power
of each member of the voting body, one often uses the original concept of
Penrose. The \textsl{a priori voting power} in his approach is proportional to
the probability that a vote cast by a given player in a hypothetical ballot
will be decisive: should this country decide to change its vote, the winning
coalition would fail to satisfy the qualified majority condition. Without any
further information about the voting body it is natural to assume that all
potential coalitions are equally likely. This very assumption leads to the
concept of \textsl{Penrose-Banzhaf index} (\textsl{PBI}) called so after John
Banzhaf (1965), who introduced this index independently.

Recent research on voting power was partially stimulated by the political
debate on the voting system used in the Council of Ministers of the European
Union (EU). The \textsl{double majority} system endorsed in 2004 by The Treaty
Establishing a Constitution for Europe, based on `per capita' and `per state'
criteria, was criticized by several authors (e.g. Paterson and Sil\'{a}rszky
(2003), Baldwin and Widgr\'{e}n (2004), Bilbao (2004), Bobay (2004), Cameron
(2004), Kirsch (2004), Plechanovov\'{a} (2004), \.{Z}yczkowski and S\l
omczy\'{n}ski (2004), Plechanovov\'{a} (2006), Taagepera and Hosli (2006),
Algaba et al. \negthinspace(2007)), who pointed out that it is favourable to
the most and to the least populated EU countries at the expense of all medium
size states. Ironically, a similar conclusion follows from a book written
fifty years earlier by Penrose, who also discovered this drawback of a `double
majority' system.\footnote{Penrose (1952) wrote: `[...] if two votings were
required for every decision, one on a \textit{per capita} basis and the other
upon the basis of a single vote for each country. This system [...] would be
inaccurate in that it would tend to favour large countries.'}

In search for an optimal two-tier voting system (where a set of constituencies
of various size elect one delegate each to a decision-making body) Penrose
(1946) considered first a direct election in a state consisting of $M$ voters
and proved that the voting power of a single citizen decays as $1/\sqrt{M}$,
provided that the votes are uncorrelated. To compensate this effect he
suggested that the a priori \textsl{voting power} of each representative in
the voting body should behave proportionally to $\sqrt{M}$ making the
citizens' voting power in all states equal and so the whole system
\textsl{representative} (the \textsl{Penrose square root law}).

Systems, where the \textsl{voting weight} of each state is proportional to the
square root of its population were discussed by several authors in the EU
context, see Laruelle and Widgr\'{e}n (1998), Hosli (2000), Tiilikainen and
Widgr\'{e}n (2000), Felsenthal and Machover (2001, 2002), Laruelle and
Valenciano (2002), Moberg (2002), Mabille (2003), Widgr\'{e}n (2003), College
of Europe (2004), Felsenthal and Machover (2004b), Hosli and Machover (2004),
Plechanovov\'{a} (2004), Widgr\'{e}n (2004). Different experts have proposed
different quotas for a square root voting systems, usually varying from 60\%
to 74\%. Clearly, the choice of an appropriate decision-taking quota
(threshold) $q$ affects both the distribution of voting power in the Council
(and thus also the representativeness of the system) and the voting system's
efficiency and transparency.

However, the assertion that the voting weight of each country should be
proportional to the square root of its population does not entirely solve the
problem of distributing the power. Kirsch (2004) expressed this as follows:
`The square root law tells us how the power should be distributed among the
countries. It is, however not clear at a first glance how to implement it in
terms of voting weights, as the voting weights do not give the power indices
immediately'. Accordingly, the question arise: how to solve the
\textsl{inverse problem}, i.e. how to allocate weights and how to set
\textsl{quota} (threshold) for qualified majority (the Council reaches a
decision when the sum of the weights of the Member States voting in favour
exceeds the threshold) to obtain required distribution of power, see Laruelle
and Widgr\'{e}n (1998), Sutter (2000), Leech (2002), Lindner and Machover
(2004), Widgr\'{e}n (2004), Pajala (2005), Aziz et al. (2007).

The answer we proposed in (\.{Z}yczkowski and S\l omczy\'{n}ski (2004)) is
surprisingly simple: one should choose the weights to be also proportional to
the square root of the population and then find such an optimal quota
$q_{\ast}$ that would produce the maximally \textsl{transparent} system, that
is a system under which the voting power of each Member State would be
approximately equal to its voting weight, or more precisely, the mean
discrepancy $\Delta$ between the voting power of each state and the rescaled
root of its population would be minimal. Then the Penrose law would be
practically fulfilled, and the potential influence of every citizen of each
Member State on the decisions taken in the Council would be almost the same.

For a concrete distribution of population in the EU consisting of 25 (resp.
27) member states it was found in (\.{Z}yczkowski and S\l omczy\'{n}ski
\negthinspace(2004), \.{Z}yczkowski et al. \negthinspace(2006), see also Feix
et al. \negthinspace(2007)) that the discrepancy exhibits a sharp minimum
around a critical quota $q_{_{{\ast}}}\sim62\,\%$ (resp. $61.5\,\%$) falling
down to a negligible value. Therefore, the Penrose square root system with
this quota is optimal, in the sense that every citizen in each member state of
the Union has the same voting power (measured by the Penrose-Banzhaf index),
i.e. the same influence on the decisions taken by the European Council. Such a
voting system occurs to give a larger voting power to the largest EU states
than the Treaty of Nice but smaller than the draft European Constitution, and
thus has been dubbed by the media as the `Jagiellonian Compromise'.

The existence of such a critical quota $q_{\ast}$ for which the rescaled PBIs
of all states are approximately equal to their voting weights, is not
restricted to this particular distribution of population in the EU. On the
contrary, it seems to be a rather generic behaviour which was found by means
of numerical simulations for typical random distributions of weights in the
voting body generated with respect to various probability measures, see
\.{Z}yczkowski and S\l omczy\'{n}ski (2004), Chang et al. \negthinspace(2006),
S\l omczy\'{n}ski and \.{Z}yczkowski (2006). The value of $q_{\ast}$ depends
to some extent on a given realization of the random population distribution,
but more importantly, it varies considerably with the number $M$ of the member
states. In the limit $M\rightarrow\infty$ the optimal quota seems to tend to
$50\%$, in consistence with the so called \textsl{Penrose limit theorem} (see
Lindner (2004), Lindner and Machover (2004)), which claims that for the quota
$50\%$ the relative power of two voters tends asymptotically to their relative
voting weight.

Working with random probability distributions it becomes difficult to get any
analytical prediction concerning the functional dependence of $q_{\ast}$ on
the number $M$ of the members of the voting body. Therefore in this work we
propose a toy model in which an analytical approach is feasible. We compute
the PBIs for this model distribution of population consisting of $M$ states
and evaluate the discrepancy $\Delta$ as a function of the quota $q$. The
optimal quota $q_{\ast}$ is defined as the value at which the quantity
$\Delta$ achieves its minimum. This reasoning performed for an arbitrary
number of states $M$ allows us to derive an explicit dependence of the optimal
quota on $M$. Results obtained analytically for this particular model occur to
be close to these received earlier in numerical experiments for random
samples. The normal approximation of the number of votes achieved by all
possible coalitions provides another estimate of the optimal quota as a
function of the quadratic mean of all the weights. The efficiency of voting
systems with optimal quota does not decrease when the number of players $M$ increases.

Applying these results we are tempted to design a simple scheme of indirect
voting (the \textsl{double square root voting systems}) based on the square
root law of Penrose supplemented by a rule setting the approximate value of
the optimal quota $q_{\ast}$ as a function either of the square root of the
number of players or the square root of the sum of their weights. Such systems
are representative, transparent and efficient.

This work is organized as follows. In Sect.~2 we recall the definition of
Penrose-Banzhaf index and define the optimal quota. Sect.~3 provides a
description of the toy model of voting in which one player is $c$ times
stronger than all other players. We describe the dependence of the optimal
quota in this model on the number of voters for $c=2$ and $c=3$. In Sect.~4 we
discuss the optimal quota applying an alternative technique of normal
approximation. The paper is concluded in Sect.~5, where we design a complete
voting system. The heuristic proof of the validity of the normal approximation
method is given in the Appendix.

\section{A priori voting power and critical quota}

Consider a set of $M$ members of the voting body, each representing a state
with population $N_{k}$, $k=1,\dots,M$. Let us denote by $w_{k}$ the voting
weight attributed to $k$-th representative. We work with renormalised
quantities, so that $\sum_{i=1}^{M}w_{i}=1$, and we assume that the decision
of the voting body is taken if the sum of the weights of all members of the
coalition exceeds the given \textsl{quota }$q\in\left[  0.5,1\right]  $, i.e.
we consider so called (\textsl{canonical}) \textsl{weighted majority voting
game} $\left[  q;w_{1},\ldots,w_{M}\right]  $, see Felsenthal and Machover (1998).

To analyse the voting power of each member one has to consider all $2^{M}$
possible coalitions and find out the number $\omega$ of winning coalitions
which satisfy the qualified majority rule adopted. The quantity $A:=\omega
/2^{M}$ measures the \textsl{decision-making efficiency} of the voting body,
i.e. the probability that it would approve a randomly selected issue. Coleman
(1971) called this quantity the \textsl{power of a collectivity to act}. For a
thorough discussion of this concept, see Lindner (2004).

The \textsl{absolute} (or \textsl{probabilistic}) \textsl{Penrose--Banzhaf
index} (\textsl{PBI}) $\psi_{k}$ of the $k$--th state is defined as the
probability that a vote cast by $k$--th representative is decisive. This
happens if $k$ is a \textsl{critical voter} in a coalition, i.e. the winning
coalition with $k$ ceases to fulfil the majority requirements without $k$.
Assuming that all $2^{M}$ coalitions are equally likely, we see that the PBI
of the $k$--th state depends only on the number $\omega_{k}$ of winning
coalitions that include this state. Namely, the number $\eta_{k}$ of
coalitions where a vote of $k$ is decisive is given by:
\begin{equation}
\eta_{k}=\omega_{k}-(\omega-\omega_{k})=2\omega_{k}-\omega\text{ .}
\label{PB1}%
\end{equation}
Moreover, the absolute Penrose-Banzhaf index of the $k$--th state is equal to
$\psi_{k}=\eta_{k}/2^{M-1}$. To compare these indices for decision bodies
consisting of different number of players, it is convenient to define the
\textsl{normalised PBIs}:
\begin{equation}
\beta_{k}:=\frac{\psi_{k}}{\sum_{i=1}^{M}\psi_{i}}=\frac{\eta_{k}}{\sum
_{i=1}^{M}\eta_{i}}\, \label{PB2}%
\end{equation}
($k=1,\dots,M$) fulfilling $\sum_{i=1}^{M}\beta_{i}=1$.

In the \textsl{Penrose voting system} one sets the voting weights proportional
to the square root of the population of each state, i.e. $w_{k}=\sqrt{N_{k}%
}/\sum_{i=1}^{M}\sqrt{N_{i}}$ for $k=1,\dots,M$. For any level of the quota
$q$ one may compute numerically the power indices $\beta_{k}$. The Penrose
rule would hold perfectly if the voting power of each state was proportional
to the square root of its population. Hence, to quantify the overall
representativeness of the voting system one can use the \textsl{mean
discrepancy} $\Delta$, defined by
\begin{equation}
\Delta:=\sqrt{\frac{1}{M}\sum_{i=1}^{M}(\beta_{i}-w_{i})^{2}}\text{ .}
\label{Delta}%
\end{equation}
The \textsl{optimal quota }$q_{\ast}$ is defined as the quota for which the
mean discrepancy $\Delta$ is minimal. Note that this quota is not unique and
usually there is a whole interval of optimal points. However, the length of
this interval decreases with increasing number of voters.

Studying the problem for a concrete distribution of population in the European
Union, as well as using a statistical approach and analyzing several random
distributions of population we found (\.{Z}yczkowski and S\l omczy\'{n}ski
(2004), \.{Z}yczkowski et al.\negthinspace\ (2006)) that in these cases all
$M$ ratios $\beta_{k}/w_{k}$ ($k=1,\dots,M$), plotted as a function of the
quota $q$, cross approximately near a single point $q_{\ast}$, i.e.
\begin{equation}
\beta_{k}\left(  q_{\ast}\right)  \approx w_{k}\left(  q_{\ast}\right)
\label{sinpoi}%
\end{equation}
for $k=1,\dots,M$. In other words, the discrepancy $\Delta$ at this
\textsl{critical quota} $q_{\ast}$ is negligible. The existence of the
critical quota was confirmed numerically in a recent study by Chang et al.
\negthinspace(2006). (This does not contradict the fact that there is a wide
range of quotas, where the mean discrepancy is relatively small, see
Widgr\'{e}n (2004), Pajala (2005).) In the next section we propose a toy model
for which a rigorous analysis of this numerical observation is possible.

\section{Toy model}

Consider a voting body of $M$ members and denote by $w_{k}$, $k=1,\dots,M$
their normalized voting weights. Assume now that a single \textsl{large}
player with weight $w_{L}:=w_{1}$ is the strongest one, while remaining
$m:=M-1$ players have equal weights $w_{S}:=w_{2}=\dots=w_{M}=(1-w_{L})/m$. We
may assume that $w_{L}\leq1/2$, since in the opposite case, for some values of
$q$, the strongest player would become a `dictator' and his relative voting
power would be equal to unity. Furthermore, we assume that the number of
\textsl{small} players $m$ is larger than two, and we introduce a parameter
$c:=w_{L}/w_{S}$ which quantifies the difference between the large player and
the other players. Thus we consider the weighted voting game $\left[
q;\frac{c}{m+c},\frac{1}{m+c},\ldots,\frac{1}{m+c}\right]  $, where the
population distribution is characterized by only two independent parameters,
say, the number of players $M$ and the ratio $c$. Sometimes it is convenient
to use as a parameter of the model the weight $w_{L}$, which is related with
the ratio $c$ by the formula $c=mw_{L}/(1-w_{L})$. On the other hand, the
qualified majority quota $q$, which determines the voting system, is treated
as a free parameter and will be optimized to minimize the discrepancy
(\ref{Delta}). Note that a similar model has been analysed in Merrill (1982).

To avoid odd-even oscillations in the discrepancy $\Delta\left(  q\right)  $
we assume that $c\geq2$. To compute the PBIs of all the players we need to
analyse three kinds of possible winning coalitions. The vote of the large
player is decisive if he forms a coalition with $k$ of his colleagues, where
$k<mq/(1-w_{L})$ and $k\geq m(q-w_{L})/(1-w_{L})$. Using the notion of the
\textsl{roof}, i.e. the smallest natural number larger than or equal to $x$,
written as $\left\lceil x\right\rceil :=\min\{n\in\mathbb{N}:n\geq x\}$, we
may put
\begin{equation}
j_{1}:=\left\lceil \frac{m(q-w_{L})}{1-w_{L}}\right\rceil -1 \label{roof3}%
\end{equation}
and
\begin{equation}
j_{2}:=\left\lceil \frac{mq}{1-w_{L}}\right\rceil -1\ \text{,} \label{roof2}%
\end{equation}
and recast the above conditions into the form
\begin{equation}
j_{1}+1\leq k\leq j_{2}\text{ .} \label{roof1}%
\end{equation}
On the other hand, there exist two cases where the vote of a small player is
decisive. He may form a coalition with $j_{2}$ other small players, or,
alternatively, he may form a coalition with the large player and $j_{1}$ small players.

With these numbers at hand, we may write down the absolute Penrose--Banzhaf
indices for both players. The a priori voting power of the larger player can
be expressed in terms of binomial symbols:
\begin{equation}
\psi_{L}:=\psi_{1}=2^{-m}\sum_{k=j_{1}+1}^{j_{2}}\binom{m}{k}\text{\ ,}
\label{PBL}%
\end{equation}
while the voting power for all the small players is equal and reads:
\begin{equation}
\psi_{S}:=\psi_{2}=\dots=\psi_{M}=2^{-m}\left[  \binom{m-1}{j_{1}}+\binom
{m-1}{j_{2}}\right]  \text{\ .} \label{PBS}%
\end{equation}
It is now straightforward to renormalize the above results according to
(\ref{PB2}) and use the normalized indices $\beta_{L}$ and $\beta_{S}$ to
write an explicit expression for the discrepancy (\ref{Delta}), which depends
on the quota $q$. Searching for an `ideal' system we want to minimize the discrepancy%

\begin{align}
\Delta(q)  &  =\frac{1}{\sqrt{M}}\sqrt{\left(  \beta_{L}-w_{L}\right)
^{2}+m\left(  \beta_{S}-w_{S}\right)  ^{2}}\nonumber\\
&  =\frac{1}{\sqrt{M}}\sqrt{\left(  \beta_{L}-w_{L}\right)  ^{2}+m\left(
\frac{1-\beta_{L}}{m}-\frac{1-w_{L}}{m}\right)  ^{2}}\\
&  =\sqrt{\frac{1+1/m}{M}}\left|  \beta_{L}-w_{L}\right| \\
&  =\frac{1}{\sqrt{m}}\left|  \beta_{L}-\frac{c}{m+c}\right| \nonumber\\
&  =\frac{1}{\sqrt{m}}\left|  \frac{\sum_{k=j_{1}+1}^{j_{2}}\binom{m}{k}}%
{\sum_{k=j_{1}+1}^{j_{2}}\binom{m}{k}+m\left(  \binom{m-1}{j_{1}}+\binom
{m-1}{j_{2}}\right)  }-\frac{c}{m+c}\right| \nonumber\\
&  =\frac{1}{\sqrt{m}}\left|  \frac{\sum_{k=\left\lceil d-c\right\rceil
}^{\left\lceil d\right\rceil -1}\binom{m}{k}}{\sum_{k=\left\lceil
d-c\right\rceil }^{\left\lceil d\right\rceil -1}\binom{m}{k}+m\left(
\binom{m-1}{\left\lceil d-c\right\rceil -1}+\binom{m-1}{\left\lceil
d\right\rceil -1}\right)  }-\frac{c}{m+c}\right|  \text{ ,} \label{DIS1}%
\end{align}
where $d:=mq/\left(  1-w_{L}\right)  =\left(  m+c\right)  q$.

In principle, one may try to solve this problem looking first for the optimal
$d$ and then computing the optimal quota $q_{\ast}$, but due to the roof in
the bounds of the sum the general case is not easy to work with.

The problem simplifies significantly if we set $c=2$, considering the
$M$--point weight vector $(w_{L},w_{L}/2,\ldots,w_{L}/2)$, where
$w_{L}=2/\left(  M+1\right)  $.

In such a case, (\ref{DIS1}) becomes
\begin{align}
\Delta(q)  &  =\frac{1}{\sqrt{m}}\left|  \frac{\binom{m}{r-2}+\binom{m}%
{r-1}\ }{\binom{m}{r-2}+\binom{m}{r-1}\ +m\left(  \binom{m-1}{r-3}+\binom
{m-1}{r-1}\right)  }-\frac{2}{m+2}\right| \nonumber\\
&  =\frac{1}{\left(  m+2\right)  \sqrt{m}}\left|  \frac{m^{2}-4mr+5m+4r^{2}%
-12r+8}{m^{2}-2mr+4m+2r^{2}-6r+5}\right|  \text{ ,} \label{DIS2}%
\end{align}
where $r:=\left\lceil d\right\rceil =\left\lceil \left(  M+1\right)
q\right\rceil $. To analyse this dependence we introduce a new variable
\begin{equation}
t:=r-M/2-1=\left\lceil \left(  M+1\right)  q\right\rceil -M/2-1\text{ ,}
\label{NEWVAR}%
\end{equation}
obtaining
\begin{align}
\Delta(t)  &  =\frac{2}{\left(  M+1\right)  \sqrt{M-1}}\frac{\left|
M-4t^{2}\right|  }{M^{2}+4t^{2}}\nonumber\\
&  =\frac{4}{\left(  M+1\right)  \sqrt{M-1}}\frac{\sqrt{M}+2t}{M^{2}+4t^{2}%
}\left|  \sqrt{M}/2-t\right|  \text{ .} \label{DIS3}%
\end{align}

In principle, one can minimize this expression finding $\min\Delta(t)=0$ for
$t_{\ast}=\sqrt{M}/2$, see Fig.1a. However, due to the presence of the roof
function in (\ref{NEWVAR}), $\Delta\left(  q\right)  $ is not a continuous
function of the quota, and, consequently, the optimization problem $\min
\Delta(q)$ does not have a unique solution and the minimal value may be
greater than $0$, see Fig.1b. Nevertheless, applying (\ref{NEWVAR}) and
(\ref{DIS3}), one can show that there exists an optimal quota $q_{\ast}\left(
M\right)  $ in the interval
\begin{equation}
\frac{M+\sqrt{M}}{2(M+1)}\leq q_{\ast}\left(  M\right)  \leq\frac{2+M+\sqrt
{M}}{2(M+1)}\ \text{.} \label{RR3}%
\end{equation}

\begin{figure}[ptbh]
\begin{center}
\ \includegraphics[width=13cm,angle=0]{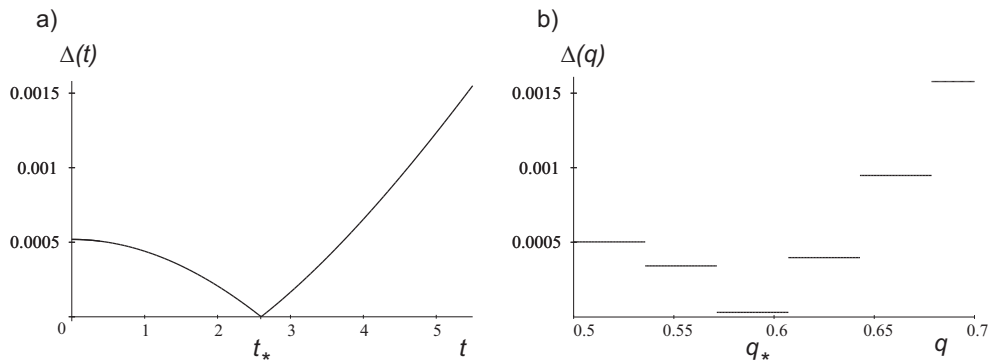}
\end{center}
\caption{a) The `mean discrepancy' $\Delta\left(  t\right)  $ as a function of
the parameter $t$; \newline b) The mean discrepancy $\Delta\left(  q\right)  $
as a function of the parameter $q$ (in both cases $c=2$, $M=27$)}%
\label{fig1}%
\end{figure}

This means that for a large number $M$ of players the optimal quota behaves
exactly as
\begin{equation}
q_{\ast}\left(  M\right)  \simeq q_{s}\left(  M\right)  :=\frac{1}{2}\left(
1+\frac{1}{\sqrt{M}}\right)  \ \text{.} \label{Ropt}%
\end{equation}
Although this is an asymptotic formula, it works also for a moderate number of
states. Moreover, it follows from (\ref{DIS2}) and (\ref{RR3}) that the
minimal mean discrepancy $\Delta(q_{\ast}\left(  M\right)  )\leq8/M^{3}$.

\bigskip Surprisingly, the efficiency of the system given by
\begin{align}
A\left(  q_{s}\left(  M\right)  \right)   &  =\frac{\sum_{k=r\left(  M\right)
-2}^{M-1}\binom{M-1}{k}+\sum_{k=r\left(  M\right)  }^{M-1}\binom{M-1}{k}%
}{2^{M}}\nonumber\\
&  =\frac{\sum_{k=r\left(  M\right)  }^{M}\binom{M}{k}+\binom{M-1}{r-2}}%
{2^{M}}\text{ ,}%
\end{align}
where $r\left(  M\right)  :=\left\lceil \left(  M+1\right)  q_{s}\left(
M\right)  \right\rceil $, does not decrease with the number of players to $0$.
On the contrary, it is always larger than $15/128\approx0.117$ and, according
to the central limit theorem, it tends to $1-\Phi(1)\approx0.159$ for
$M\rightarrow\infty$.

Analogous considerations for $c=3$ give similar result:
\begin{equation}
\frac{1+M+\sqrt{M}}{2(M+1)}\leq q_{\ast}\left(  M\right)  \leq\frac
{3+M+\sqrt{M}}{2(M+1)}\text{ ,} \label{TM3}%
\end{equation}
and so, also in this case, $q_{\ast}\left(  M\right)  \simeq\frac{1}%
{2}(1+1/\sqrt{M})$.

\section{Normal approximation}

Let us have a closer look at the approximate formula (\ref{Ropt}) for the
optimal quota. In the limit $M\rightarrow\infty$ the optimal quota tends to
$1/2$ in agreement with the Penrose limit theorem, see Lindner (2004), Lindner
and Machover (2004). Numerical values of the approximate optimal quota $q_{s}$
obtained in our toy model for $c=2$ and $c=3$ are consistent, with an accuracy
up to two per cent, with the data obtained numerically by averaging quotas
over a sample of random weights distributions (generated with respect to the
statistical measure, i.e. the symmetric Dirichlet distribution with Jeffreys'
priors), see \.{Z}yczkowski and S\l omczy\'{n}ski (2004), \.{Z}yczkowski et
al. \negthinspace(2006).\footnote{Nevertheless, one can construct an
artificial model with different values of optimal quota. In this aim, it is
enough to consider one `small' state and an even number of `large' states with
equal population (i.e. $c<1$ in our toy model), see Lindner (2004), Lindner
and Machover (2004). As Lindner stressed: `experience suggests that such
counter-examples are atypical, contrived exceptions'.} Furthermore, the above
results belong to the range of values of the quota for qualified majority,
which have been used in practice or recommended by experts on designing the
voting systems.

Consider now a voting body of $M$ members and denote by $w_{k}$, $k=1,\dots
,M$, their normalized voting weights fulfilling $\sum_{i=1}^{M}w_{i}=1$. Feix
et al. \negthinspace(2007)\textit{\ }proposed (also in the EU context) yet
another method of estimating the optimal quota for the weighted voting game
$\left[  q;w_{1},\ldots,w_{M}\right]  $, where $q\in\left[  0.5,1\right]  $.
They considered the histogram $n$ of the sum of weights (number of votes)
achieved by all possible coalitions
\begin{equation}
n\left(  z\right)  =\frac{\operatorname*{card}\left\{  I\subset\left\{
1,\ldots,M\right\}  :\sum_{i\in I}w_{i}=z\right\}  }{2^{M}} \label{hist1}%
\end{equation}
and assumed that it allows the normal approximation with the mean value
$m=\frac{1}{2}\sum_{i=1}^{M}w_{i}=\frac{1}{2}$ and the variance $\sigma
^{2}=\frac{1}{4}\sum_{i=1}^{M}w_{i}^{2}$, i.e.
\begin{equation}
\mathcal{N}\left(  q\right)  :=\sum_{z\leq q}n\left(  z\right)  \approx
\int_{-\infty}^{q}\frac{1}{\sigma\sqrt{2\pi}}\exp\left(  -\frac{\left(
x-m\right)  ^{2}}{2\sigma^{2}}\right)  dx=\Phi\left(  \frac{q-m}{\sigma
}\right)  \text{ ,} \label{hist2}%
\end{equation}
where $\Phi$ is the standard normal cumulative distribution function. The
authors argued that for the quota close to the inflection point $q_{n}%
:=m+\sigma$ of the normal curve, where the `density' of the histogram is
approximately linear, the ratios $\beta_{k}/w_{k}$ ($k=1,\dots,M$) are close
to $1$. In other words, the quota $q_{n}$ is close to the optimal quota
$q_{\ast}$.\ In Appendix we show how this fact follows from the normal
approximation formula for the absolute Banzhaf indices. In particular we use
heuristic arguments to demonstrate that in this case
\begin{equation}
\psi_{k}\approx\sqrt{\frac{2}{\pi e}}\frac{w_{k}}{\sqrt{\sum_{i=1}^{M}%
w_{i}^{2}}} \label{absBanapp}%
\end{equation}
and, in consequence,
\begin{equation}
\beta_{k}\approx w_{k} \label{norBanapp}%
\end{equation}
for $k=1,\dots,M$. The validity of this method depends on the accuracy of the
normal approximation for the absolute Banzhaf indices (see Appendix). The
necessary condition for the latter is
\begin{equation}
\max_{j=1,\ldots,M}w_{j}\ll\sqrt{\sum_{i=1}^{M}w_{i}^{2}}\text{ .}
\label{neccon}%
\end{equation}
For the thorough discussion of the problem see Owen (1975), Leech (2003),
Lindner (2004), Feix et al. \negthinspace(2007). For the Penrose voting
system, where $w_{k}\sim\sqrt{N_{k}}$ ($k=1,\ldots,M$), (\ref{neccon}) is
equivalent to
\begin{equation}
\max_{j=1,\ldots,M}N_{j}\ll\sum_{i=1}^{M}N_{i}\text{ ,} \label{necconpop}%
\end{equation}
which means that the population of each country is relatively small when
compared with the total population of all countries. One can easily check that
it is more likely that (\ref{neccon}) holds in this case than when the weights
are proportional to the population.

Approximating the optimal quota $q_{\ast}$ by the inflection point of the
normal distribution, $q_{n}=m+\sigma$, we arrive at an explicit
weights-dependent formula for the optimal quota:
\begin{equation}
q_{\ast}\simeq q_{n}\left(  w_{1},\ldots,w_{M}\right)  :=m+\sigma=\frac{1}%
{2}\left(  1+\sqrt{\sum_{i=1}^{M}w_{i}^{2}}\,\right)  \text{ .}
\label{NORQOUTA}%
\end{equation}
This approximation of the optimal quota can be directly compared with the
approximation (\ref{Ropt}) obtained for the toy model. Since $\sum_{i=1}%
^{M}w_{i}=1$ implies $\sum_{i=1}^{M}w_{i}^{2}\geq1/M$, it follows that
\begin{equation}
q_{s}\left(  M\right)  =\frac{1}{2}\left(  1+\frac{1}{\sqrt{M}}\right)
\leq\frac{1}{2}\left(  1+\frac{1}{\sqrt{M_{eff}}}\right)  =q_{n}\text{ ,}
\label{INEQ}%
\end{equation}
where $M_{eff}:=1/\sum_{i=1}^{M}w_{i}^{2}$ is equal to the \textsl{effective
number of players}. (This quantity was introduced by Laakso and Taagepera
(1979) and is the inverse of the more widely used
\textsl{Herfindahl--Hirschman index of concentration} (Hirschman (1945),
Herfindahl (1950), see also Feld and Grofman (2007).) The equality in
(\ref{INEQ}) holds if and only if all the weights are equal. For the Penrose
voting system we have
\begin{equation}
q_{n}=\frac{1}{2}\left(  1+\frac{\sqrt{\sum_{i=1}^{M}N_{i}}}{\sum_{i=1}%
^{M}\sqrt{N_{i}}}\right)  \text{ ,} \label{norquoPen}%
\end{equation}
where $N_{k}$ stands for the population of the $k$-th country. For the toy
model we get $q_{n}=\frac{1}{2}\left(  1+\frac{\sqrt{M+c^{2}-1}}%
{M+c-1}\right)  \simeq q_{s}\left(  M\right)  $ for large$~M$.

Both approximations $q_{s}$ and $q_{n}$ are consistent with an accuracy up to
two per cent, with the optimal quotas $q_{\ast}$ obtained for the Penrose
voting system applied retrospectively to the European Union (see Tab.~1
below). Observe that in this case the approximation of the optimal quota
$q_{\ast}$ by $q_{n}$ is better for larger number of states, where the normal
approximation of the histogram is more efficient.\medskip

\begin{center}%
\begin{tabular}
[c]{||c|c|c|c||}\hline\hline
$M$ & $15$ & $25$ & $27$\\\hline
year & $1995$ & $2004$ & $2007$\\\hline
$q_{s}\left[  \%\right]  $ & $62.9$ & $60.0$ & $59.6$\\\hline
$q_{\ast}\left[  \%\right]  $ & $64.4$ & $62.0$ & $61.5$\\\hline
$q_{n}\left[  \%\right]  $ & $64.9$ & $62.2$ & $61.6$\\\hline\hline
\end{tabular}
\bigskip

Tab. 1. Comparison of optimal quotas for the Penrose voting system applied to
the EU ($q_{\ast}$) and for two approximations ($q_{s}$, $q_{n}$%
).\footnote{The calculations are based on data from: \textsl{50 years of
figures on Europe. Data 1952-2001.} Office for Official Publications of the
European Communities: Luxembourg; 2003, and on data from: EUROSTAT: Lanzieri
G. Population in Europe 2005: first results. Statistics in focus. Population
and social conditions 2006; 16: 1-12.}\bigskip
\end{center}

Applying the normal approximation one can easily explain why the efficiency
$A$ of our system does not decrease when the number of players $M$ increases.
We have
\begin{equation}
A(q_{s})\geq A\left(  q_{n}\right)  \approx1-\mathcal{N}\left(  q_{n}\right)
\approx\int_{m+\sigma}^{\infty}\frac{1}{\sigma\sqrt{2\pi}}\exp\left(
-\frac{\left(  x-m\right)  ^{2}}{2\sigma^{2}}\right)  dx\text{ .} \label{EFF2}%
\end{equation}
The right-hand side of this inequality depends neither on $m$ nor on $\sigma$,
and it equals $1-\Phi(1)\approx15.9 \% $, where $\Phi$ is the standard normal
cumulative distribution function.

\section{Double square root voting system}

We shall conclude this paper proposing a complete voting system based on the
Penrose square root law. The system consists of a single criterion only and is
determined by the following two rules:

\medskip\textbf{A.} The voting weight attributed to each member of the voting
body of size $M$ is proportional to the square root of the population he or
she represents;

\medskip\textbf{B.} The decision of the voting body is taken if the sum of the
weights of members of a coalition exceeds the quota $q_{s}=(1+1/\sqrt{M})/2$.\medskip

These rules characterize the \textsl{double square root} system: On one hand,
the weight of each state is proportional to the square root of its population,
on the other hand, the quota decreases to $0.5$ inversely proportionally to
the square root of the size of the voting body. If the weights $w_{i}$ are
fixed, one can set the quota to $q_{n}=(1+(\sum_{i=1}^{M}w_{i}^{2})^{1/2})/2$,
or just take the optimal quota $q_{\ast}$ which, however, requires more
computational effort.

Such a voting system is extremely simple, since it is based on a single
criterion. It is objective and so cannot a priori handicap a given member of
the voting body. The quota for qualified majority is considerably larger than
$50\%$ for any size of the voting body of a practical interest. Thus the
voting system is also moderately conservative. Furthermore, the system is
representative and transparent: the voting power of each member of the voting
body is (approximately) proportional to its voting weight. However, as a
crucial advantage of the proposed voting system we would like to emphasize its
extendibility: if the size $M$ of the voting body changes, all one needs to do
is to set the voting weights according to the square root law and adjust the
quota. The system is also moderately efficient: as the number $M$ grows, the
efficiency of the system does not decrease.

The formulae for the quotas $q_{s}\left(  M\right)  $ and $q_{n}$ can be also
applied in other weighted voting games. Note that for a fixed number of
players the quota $q_{s}\left(  M\right)  $ does not depend on the particular
distribution of weights in the voting body. This feature may be relevant, e.g.
for voting bodies in stock companies where the voting weights of stockholders
depend on the proportion of stock that investors hold and may vary frequently.

Although the limiting behaviour $M\rightarrow\infty$ may not necessarily be
interesting for politicians, our work seems to have some practical
implications for the on-going debate concerning the voting system in the
Council of the EU. Since the number of Member States is not going to be
explicitly provided in the text of the European Constitution, one should
rather avoid to include any specific threshold for the qualified majority. In
fact the optimal quota depends on the number of members of the voting body, so
there should be a possibility to adjust it in future without modifying the
European Constitution.\bigskip\pagebreak 

\begin{center}
{\large Acknowledgements\medskip}
\end{center}

It is a pleasure to thank W.~Kirsch, A.~Pajala, T.~Soza{\'{n}}ski, and
T.~Zastawniak for fruitful discussions, and to F.~Bobay, M.~Machover,
A.~Moberg, and E.~Ratzer for beneficial correspondence on voting systems. We
also would like to thank an anonymous referee for helping us to clarify
several important points of our reasoning. This work was supported in part by
the Marie Curie Actions Transfer of Knowledge project COCOS (contract
MTKD-CT-2004-517186) and the MNiSW grant no 1 H02E 052 30.\medskip

\textbf{Note added. }After this work was completed we learned about a related
work by Lindner and Owen (2007), in which the same toy model was investigated.\bigskip

\begin{center}
{\large Appendix: Optimal quota for the normal approximation\medskip}
\end{center}

In this appendix we show that in vicinity of the inflection point
$q_{n}=m+\sigma$ of the normal distribution the relative Banzhaf indices
$\beta_{j}$ are close to the weights $w_{j}$. This reasoning holds in
particular for the Penrose voting system, for which the weights are
proportional to the square root of the populations.

\medskip

Consider a weighted voting game $\left[  q;w_{1},\ldots,w_{M}\right]
\medskip$, where $q\in\left[  0.5,1\right]  $ and $\sum_{i=1}^{M}w_{i}%
=1$.\medskip\ Set $m:=\frac{1}{2}\sum_{i=1}^{M}w_{i}=\frac{1}{2}$ and
$\sigma^{2}:=\frac{1}{4}\sum_{i=1}^{M}w_{i}^{2}$. Let $j=1,\ldots,M$. We put
$m_{j}:=m-w_{j}/2$ and $\sigma_{j}^{2}:=\sigma^{2}-w_{j}^{2}/4$.\smallskip

The absolute Banzhaf index
\begin{equation}
\psi_{j}=\Pr\left(  \left\{  I\subset\left\{  1,\ldots,M\right\}  :q-w_{j}%
\leq\sum_{i\in I,i\neq j}w_{i}<q\right\}  \right)  \tag{A1}\label{Ban}%
\end{equation}
is equal to the difference of the number of wining coalitions formed with and
without the $j$--th player divided by $2^{M-1}$. A key step in our reasoning
is to assume that the sum of weights of the members of a coalition can be
approximated by the normal distribution. This assumption implies that the
Banzhaf index $\psi_{j}$ is approximately equal to the difference of two
normal cumulative distribution functions taken at two points shifted by the
corresponding weight $w_{j}$,
\begin{equation}
\psi_{j}\approx\Phi\left(  q;m_{j},\sigma_{j}\right)  -\Phi\left(
q-w_{j};m_{j},\sigma_{j}\right)  \text{ .} \tag{A2}\label{Banapp1}%
\end{equation}
Here $\Phi\left(  x;\mu,d\right)  =\Phi\left(  (x-\mu)/d\right)  $ stands for
the normal cumulative distribution function with mean $\mu$ and standard
deviation $d$. Therefore
\begin{equation}
\psi_{j}\approx\Phi\left(  \frac{q-m_{j}}{\sigma_{j}}\right)  -\Phi\left(
\frac{q-w_{j}-m_{j}}{\sigma_{j}}\right)  \text{ .} \tag{A3}\label{Banapp2}%
\end{equation}

We are going to analyse the behaviour of the power indices at the inflection
point, $q=q_{n}:=m+\sigma$. In such a case,
\begin{align}
\psi_{j}  &  \approx\Phi\left(  \frac{m+\sigma-m_{j}}{\sigma_{j}}\right)
-\Phi\left(  \frac{m+\sigma-w_{j}-m_{j}}{\sigma_{j}}\right) \nonumber\\
&  =\Phi\left(  \frac{\sigma+\frac{1}{2}w_{j}}{\sigma_{j}}\right)
-\Phi\left(  \frac{\sigma-\frac{1}{2}w_{j}}{\sigma_{j}}\right)  \smallskip
\nonumber\\
&  =\Phi\left(  \sqrt{\frac{1+v_{j}}{1-v_{j}}}\,\right)  -\Phi\left(
\sqrt{\frac{1-v_{j}}{1+v_{j}}}\,\right)  \text{ ,} \tag{A4}\label{Banapp3}%
\end{align}
where $v_{j}:=w_{j}/2\sigma=w_{j}/\sqrt{\sum_{i=1}^{M}w_{i}^{2}}$. If
$w_{j}\ll\sqrt{\sum_{i=1}^{M}w_{i}^{2}}$, then $v_{j}\ll1$, and both
$\sqrt{\left(  1+v_{j}\right)  /\left(  1-v_{j}\right)  }$ and $\sqrt{\left(
1-v_{j}\right)  /\left(  1+v_{j}\right)  }$ are close to $1$. Near this point
the standard normal density function $\Phi^{\prime}$ is almost linear and so
\begin{equation}
\Phi\left(  \frac{\sigma+\frac{1}{2}w_{j}}{\sigma_{j}}\right)  -\Phi\left(
\frac{\sigma-\frac{1}{2}w_{j}}{\sigma_{j}}\right)  \approx\Phi^{\prime}\left(
\frac{\sigma}{\sigma_{j}}\right)  \frac{w_{j}}{\sigma_{j}}\text{ .}
\tag{A5}\label{Norapp}%
\end{equation}
>From (\ref{Banapp3}) and (\ref{Norapp}) we deduce that
\begin{align}
\psi_{j}  &  \approx\Phi^{\prime}\left(  \frac{\sigma}{\sigma_{j}}\right)
\frac{w_{j}}{\sigma_{j}}\nonumber\\
&  =\frac{1}{\sqrt{2\pi}}\frac{w_{j}}{\sigma_{j}}\exp\left(  -\frac{\sigma
^{2}}{2\sigma_{j}^{2}}\right) \nonumber\\
&  =\sqrt{\frac{2}{\pi}}\frac{w_{j}}{\sqrt{\left(  \sum_{i=1}^{M}w_{i}%
^{2}\right)  -w_{j}^{2}}}\exp\left(  -\frac{\sum_{i=1}^{M}w_{i}^{2}}{2\left(
\left(  \sum_{i=1}^{M}w_{i}^{2}\right)  -w_{j}^{2}\right)  }\right)
\nonumber\\
&  =\sqrt{\frac{2}{\pi}}\frac{v_{j}}{\sqrt{1-v_{j}^{2}}}\exp\left(  -\frac
{1}{2\left(  1-v_{j}^{2}\right)  }\right) \nonumber\\
&  =\sqrt{\frac{2}{\pi e}}v_{j}+o\left(  v_{j}^{4}\right)  \text{ .}
\tag{A6}\label{Banapp4}%
\end{align}
Consequently,
\begin{equation}
\psi_{j}\approx\sqrt{\frac{2}{\pi e}}\frac{w_{j}}{\sqrt{\sum_{i=1}^{M}%
w_{i}^{2}}}+o\left(  v_{j}^{4}\right)  \text{ ,} \tag{A7}\label{Banapp5}%
\end{equation}
and so
\begin{equation}
\frac{\beta_{j}}{w_{j}}\approx1\text{ .} \tag{A8}\label{norBanapp1}%
\end{equation}

The above reasoning shows that if we select the inflection point $q_{n}$ for
the quota $q$, then all the normalised Penrose--Banzhaf indices $\beta_{j}$
are approximately equal to the weights $w_{j}$, and so $q_{n}$ must be close
to the critical quota $q_{\ast}$ The accuracy of this approximation depends
highly on the accuracy of the normal approximation in (\ref{Banapp1}).

Note that for the quota $q=m=1/2$ we get (see Lindner (2004), Lindner and
Machover (2004) for the formal proof)
\begin{align}
\psi_{j}  &  \approx\Phi^{\prime}\left(  0\right)  \frac{w_{j}}{\sigma_{j}%
}\nonumber\\
&  =\sqrt{\frac{2}{\pi}}\frac{w_{j}}{\sqrt{\left(  \sum_{i=1}^{M}w_{i}%
^{2}\right)  -w_{j}^{2}}}\nonumber\\
&  =\sqrt{\frac{2}{\pi}}\frac{w_{j}}{\sqrt{\sum_{i=1}^{M}w_{i}^{2}}}+o\left(
v_{j}^{2}\right)  \text{ .} \tag{A9}\label{symBanapp1}%
\end{align}
In this case, the second order terms in $v_{j}$ are present and, in
consequence, the indices $\beta_{j}$ need not be as close to $w_{j}$ as for
$q=q_{n}$, where (\ref{Banapp5}) holds up to corrections of order four in
$v_{j}$. Moreover, it is interesting to note, that increasing the threshold
from $m=1/2$ to $q_{n}=m+\sigma=1/2+\sigma$ causes the decrease of the
absolute power indices $\phi_{j}$ by the factor $1/\sqrt{e}\approx0.607$.\pagebreak 

\textbf{References}\bigskip

Algaba, E., Bilbao, J.M. and Fern\'{a}ndez, J.R. (2007), The distribution of
power in the European Constitution, \textit{European Journal of Operational
Research} 176: 1752-1766.\smallskip

Aziz, H., Paterson, M. and Leech, D. (2007), \textit{Efficient Algorithm for
Designing Weighted Voting Games}. Preprint. \newline
{\small http://www.dcs.warwick.ac.uk/reports/cs-rr-434.pdf}\smallskip

Baldwin, R.E. and Widgr\'{e}n, M. (2004), \textit{Council Voting in the
Constitutional Treaty: Devil in the Details}. (CEPS Policy Briefs No. 53;
Centre for European Policy Studies, Brussels) \newline
{\small http://hei.unige.ch/\%7Ebaldwin/PapersBooks/Devil\_in\_the\_details\_\allowbreak
BaldwinWidgren.pdf}\smallskip

Banzhaf, J.F. (1965), Weighted voting does not work: A mathematical analysis,
\textit{Rutgers Law Review} 19: 317-343.\smallskip

Bilbao, J.M. (2004), \textit{Voting Power in the European Constitution}.
Preprint.\newline {\small http://www.esi2.us.es/\symbol{126}%
mbilbao/pdffiles/Constitution.pdf}\smallskip

Bobay, F. (2004), Constitution europ\'{e}enne: redistribution du pouvoir des
\'{E}tats au Conseil de l'UE, \textit{\'{E}conomie et Pr\'{e}vision} 163: 101-115.\smallskip

Cameron, D.R. (2004), The stalemate in the constitutional IGC over the
definition of a qualified majority, \textit{European Union Politics} 5: 373-391.\smallskip

Chang, P.-L., Chua, V.C.H. and Machover, M. (2006), L S Penrose's limit
theorem: Tests by simulation, \textit{Mathematical Social Sciences} 51: 90-106.\smallskip

Coleman, J.S. (1971), Control of collectivities and the power of a
collectivity to act, in: B. Lieberman (ed.), \textit{Social Choice}, New York:
Gordon and Breach. Reprinted in: J.S. Coleman, 1986, \textit{Individual
Interests and Collective Action}, Cambridge University Press.\smallskip

College of Europe (2004), \textit{Making Europe Work: A Compromise Proposal on
Voting in the Council}. College of Europe: Warsaw and Bruges. \newline
{\small http://www.coleurop.be/content/publications/pdf/\allowbreak
MakingEuropeWork.pdf}\smallskip

Feix, M.R., Lepelley, D., Merlin, V. and Rouet, J.L. (2007), On the voting
power of an alliance and the subsequent power of its members, \textit{Social
Choice and Welfare} 28: 181-207.\smallskip

Feld, S.L. and Grofman, B. (2007), The Laakso-Taagepera index in a mean and
variance framework, \textit{Journal of Theoretical Politics} 19: 101-106.\smallskip

Felsenthal, D.S. and Machover, M. (1998), \textsl{Measurement of Voting Power:
Theory and Practice, Problems and Paradoxes}. Edward Elgar: Cheltenham.\smallskip

Felsenthal, D.S. and Machover, M. (2001), Treaty of Nice and qualified
majority voting, \textit{Social Choice and Welfare} 18: 431-464.\smallskip

Felsenthal, D.S. and Machover, M. (2002), \textit{Enlargement of the EU and
Weighted Voting in its Council of Ministers }[online]. LSE Research Online:
London.\newline {\small http://eprints.lse.ac.uk/407/01/euenbook.pdf}\smallskip

Felsenthal, D.S. and Machover M. (2004a), A priori voting power: What is it
all about? \textit{Political Studies Review} 2: 1-23.\smallskip

Felsenthal, D.S. and Machover M. (2004b), Qualified majority voting explained,
\textit{Homo Oeconomicus} 21: 573-576.\smallskip

Herfindahl, O.C. (1950), \textit{Concentration in the Steel Industry}. PhD
Dissertation; Columbia University.\smallskip

Hirschman, A.O. (1945), \textsl{National Power and Structure of Foreign
Trade}. University of California Press: Berkeley.\smallskip

Hosli, M.O. (2000), \textit{Smaller States and the New Voting Weights in the
Council}. (Working Paper, Netherlands Institute of International Relations,
Clingendael, July 2000)\newline
{\small http://www.clingendael.nl/publications/2000/20000700\_cli\_ess\_hosli.pdf}%
\smallskip

Hosli, M.O. and Machover, M. (2004), The Nice Treaty and voting rules in the
Council: a reply to Moberg (2002), \textit{Journal of Common Market Studies}
42: 497-521.\smallskip

Kirsch, W. (2004), \textit{The New Qualified Majority in the Council of the
EU. Some Comments on the Decisions of the Brussels Summit}. Preprint.\newline
{\small http://www.ruhr-uni-bochum.de/mathphys/politik/eu/Brussels.pdf}\smallskip

Laakso, M. and Taagepera, R. (1979), Effective number of parties: a measure
with application to West Europe, \textit{Comparative Political Studies} 12: 3-27.\smallskip

Laruelle, A. and Valenciano, F. (2002), Inequality among EU citizens in the
EU's Council decision procedure, \textit{European Journal of Political
Economy} 18: 475-498.\smallskip

Laruelle, A. and Widgr\'{e}n, M. (1998), Is the allocation of voting power
among the EU states fair? \textit{Public Choice} 94: 317-339.\smallskip

Leech, D. (2002), Designing the voting system for the Council of the EU.
\textit{Public Choice} 113: 437-464.\smallskip

Leech, D. (2003), Computing power indices for large voting games,
\textit{Management Science} 49: 831-837.\smallskip

Lindner, I. (2004), \textit{Power Measures in Large Weighted Voting Games
Asymptotic Properties and Numerical Methods}. PhD Dissertation; Universit\"{a}%
t Hamburg. \newline
{\small http://deposit.ddb.de/cgi-bin/dokserv?idn=972400516}\smallskip

Lindner, I. and Machover, M. (2004), L. S. Penrose's limit theorem: proof of
some special cases, \textit{Mathematical Social Sciences} 47: 37-49.\smallskip

Lindner, I. and Owen, G. (2007), Cases where the Penrose limit theorem does
not hold, \textit{Mathematical Social Sciences} 53: 232-238.\smallskip

Mabille, L. (2003), \textit{Essai sur une juste pond\'{e}ration des voix au
Conseil de l'Union europ\'{e}enne}. Preprint; see also: Dubois, N.,
Pond\'{e}ration des voix: la preuve par ``27'', \textit{Liberation}
26/11/2003.\newline {\small http://pageperso.aol.fr/lcmabille/}\smallskip

Merrill III, S. (1982), Approximations to the Banzhaf index of voting power,
\textit{American Mathematical Monthly} 89: 108-110.\smallskip

Moberg, A. (2002), The Nice Treaty and voting rules in the Council,
\textit{Journal of Common Market Studies} 40: 259-82.\smallskip

Owen, G. (1975), Multilinear extensions and the Banzhaf value, \textit{Naval
Research Logistics Quarterly} 22: 741-750.\smallskip

Pajala, A. (2005), \textit{Maximal Proportionality between Votes and Voting
Power: The Case of the Council of the European Union }[online]. LSE Research
Online: London.\newline
{\small http://www.lse.ac.uk/collections/VPP/VPPpdf/VPPpdf\_Wshop4/pajala.pdf}\smallskip

Paterson, I. and Sil\'{a}rszky, P. (2003), \textit{Draft Constitution and the
IGC: Voting can be Simple and Efficient - without introducing the Massive
Transfer of Power implied by the Convention's Double Majority Proposal}.
(Institute for Advanced Studies, Vienna, December 2003).\newline
{\small http://www.ihs.ac.at/publications/lib/forum2ihsdec2003.pdf}\smallskip

Penrose, L.S. (1946), The elementary statistics of majority voting,
\textit{Journal of the Royal Statistical Society} 109: 53-57.\smallskip

Penrose, L.S. (1952), \textsl{On the Objective Study of Crowd Behaviour}. H.K.
Lewis \& Co: London.\smallskip

Plechanovov\'{a}, B. (2004), \textit{Draft Constitution and the
Decision-Making Rule for the Council of Ministers of the EU - Looking for
Alternative Solution}. (European Integration online Papers (EIoP), Vol. 8, No.
12) \newline {\small http://eiop.or.at/eiop/pdf/2004-012.pdf}\smallskip

Plechanovov\'{a}, B. (2006), Je rozhodovac\'{i} procedura v Rad\v{e}
Evropsk\'{e} unie spravedliv\'{a}? \textit{Mezin\'{a}rodn\'{i} vztahy} 1: 5-22.\smallskip

S\l omczy\'{n}ski, W. and \.{Z}yczkowski, K. (2006), Penrose voting system and
optimal quota, \textit{Acta Physica Polonica} 37: 3133-3143.\smallskip

Sutter, M. (2000), Fair allocation and re-weighting of votes and voting power
in the EU before and after the next enlargement, \textit{Journal of
Theoretical Politics} 12: 433-449.\smallskip

Taagepera, R. and Hosli, M.O. (2006), National representation in international
organizations: the seat allocation model implicit in the European Union
Council and Parliament, \textit{Political Studies} 54: 370-398.\smallskip

Tiilikainen, T. and Widgr\'{e}n, M. (2000), Decision-making in the EU: a small
country perspective, \textit{The Finnish Economy and Society} 4: 63-71.\smallskip

Widgr\'{e}n, M. (2003), \textit{Power in the Design of Constitutional Rules}.
(European Economy Group, University of Madrid Working Papers No. 23)
\newline
{\small http://www.ucm.es/info/econeuro/documentos/documentos/\allowbreak
dt232003.pdf}\smallskip

Widgr\'{e}n, M. (2004), Enlargements and the Principles of Designing EU
Decision-Making Procedures, in: Blankart, C.B. and Mueller, D.C. (eds.),
\textit{A Constitution for the European Union}, MIT Press, pp. 85-108.\smallskip

\.{Z}yczkowski, K. and S\l omczy\'{n}ski, W. (2004), \textit{Voting in the
European Union: The Square Root System of Penrose and a Critical Point.}
Preprint cond-mat.0405396; May 2004. \newline
{\small http://arxiv.org/ftp/cond-mat/papers/0405/0405396.pdf}\smallskip

\.{Z}yczkowski, K., S\l omczy\'{n}ski, W. and Zastawniak, T. (2006), Physics
for fairer voting, \textit{Physics World} 19: 35-37.

\pagebreak 
\end{document}